# Absence of charge-density-wave sliding in epitaxial charge-ordered Pr$_{0.48}$Ca$_{0.52}$MnO$_3$ films

B. Fisher[1*], J. Genossar[1], L. Patlagan[1], S. Kar-Narayan[2], X. Moya[2], D. Sánchez[2],

P. A. Midgley[2] and N. D. Mathur[2**]

* phr06bf@physics.technion.ac.il
** ndm12@cam.ac.uk

[1] *Department of Physics, Technion, Haifa 32000, Israel*
[2] *Department of Materials Science, University of Cambridge, Pembroke Street, Cambridge CB2 3QZ, UK*

For an epitaxial Pr$_{0.48}$Ca$_{0.52}$MnO$_3$ film on NdGaO$_3$, we use transmission electron microscopy to observe a "charge-ordered" superlattice along the in-plane direction *a*. The same film shows no electrical signatures of charge order. The in-plane electrical anisotropy $\rho_a/\rho_c = 28$ is constant, and there is no evidence of sliding charge density waves up to the large field of ~10$^3$ V/cm.

Perovskite manganites[1] of the form (RE$^{3+}$,AE$^{2+}$)MnO$_3$ (RE = rare earth, AE = alkaline earth) show two predominant low-temperature phases, namely the ferromagnetic metal which is highly spin polarized, and the so-called "charge-ordered" (CO) insulator in which a superlattice is experimentally observed. The CO phase was originally described in terms of Mn$^{3+}$/Mn$^{4+}$ ordering[2-5], but widespread doubt about this interpretation has arisen[6-10]. It is now accepted that this picture is inaccurate, especially given that the superlattice is slightly misoriented and locally incommensurate for[10] La$_{0.48}$Ca$_{0.52}$MnO$_3$ and[11] Pr$_{0.48}$Ca$_{0.52}$MnO$_3$. Complete charge disproportionation is coulombically expensive and a charge density wave (CDW) picture has been proposed[12-17,10]. It is therefore interesting to investigate whether it is possible to observe CDW sliding.

Recently it was argued[18] that epitaxial La$_{0.5}$Ca$_{0.5}$MnO$_3$ films on NdGaO$_3$ (001) show CDW sliding in fields of ~10$^2$ V/cm, but the measurement geometry was irregular and so inhomogeneous heating currents cannot be ruled out[19]. These measurements are difficult to replicate, as La$_{0.5}$Ca$_{0.5}$MnO$_3$ superlattice reflections in transmission electron microscopy (TEM) diffraction patterns are normally absent[20], and if present they are weak[18,21]. Strong signatures of CO have been observed in (011)-oriented films of[22,23] Nd$_{0.5}$Sr$_{0.5}$MnO$_3$ and[24] Bi$_{0.4}$Ca$_{0.6}$MnO$_3$, but this orientation promotes both a rough surface and also defects that would pin CDWs. Here we investigate an (001)-oriented epitaxial Pr$_{0.48}$Ca$_{0.52}$MnO$_3$ film in which we observe distinct superlattice reflections. The electrical resistivity shows no sign of the CO transition at

$T_{CO}$, and there is no evidence for CDW sliding. Non-Ohmic behaviour arises only at high fields (~$10^3$ V/cm) and we show this is due to heating.

A $Pr_{0.48}Ca_{0.52}MnO_3$ film was grown on an $NdGaO_3$ (001) substrate (NGO) by pulsed laser deposition ($\lambda$=248 nm, 1 Hz, 2 J cm$^{-2}$, target-substrate distance=8 cm, 800 °C, 15 Pa flowing $O_2$) from a commercial target (Praxair, USA). After deposition, the film was annealed in ~50 kPa $O_2$ for 1 hour at the growth temperature, and then cooled to room temperature. A High Resolution Philips PW3050/65 X'Pert PRO horizontal diffractometer ($CuK_{\alpha1}$) was used to confirm crystallinity, in-plane orientation and thickness (53±3 nm from interference fringes near the (004) substrate peak). Atomic force microscopy indicated a surface roughness of ~1 nm. Magnetic data could not be collected because the paramagnetic $NdGaO_3$ substrate masked the weak film signal, even when a part of the substrate was thinned to ~200 μm.

Part of the substrate was pre-thinned to ~50 μm by mechanical polishing, and then a 6 μm × 4 μm electron-transparent window was defined with a Focused Ion Beam for TEM studies. This window was ~150 nm thick such that ~100 nm of substrate remained. A Philips CM300 FEG TEM, equipped with a Gatan Image Filter and CCD camera, was used to collect energy-filtered 100 nm [010] SAD patterns. The microscope supports Gatan sample stages with base temperatures ~9.6 K (helium) and ~90 K (nitrogen).

Another part of the film (3×3 mm$^2$) was used for electrical-transport studies. Four-probe d.c. measurements and pulsed I-V characteristics were measured as described in Ref. 25. First, measurements along *a* were made by painting four silver-dag strips along *c*. Second, measurements along *c* were made by washing off the first set of strips with acetone, and painting new strips in the perpendicular direction. Third, measurements were made once more along *a* by washing off the new strips and painting strips in the original orientation. We found *a*-axis film resistivity $\rho_a(T)$ measured in the first and third rounds to be identical within a small constant factor, i.e. washing and painting does not corrupt measurements. In this paper, we present data from rounds two and three, where probes 1-4 were unevenly spaced such that the separation of the inner probes ($d_{23}$ = 0.8±0.05 mm) was more than double the separation of each outer probe and its nearest neighbour ($d_{12} \approx d_{34}$). During high-field measurements, the voltage-drop ratio $V_{14}/V_{23}$ was practically constant and scaled with the inter-probe distances $d_{23}$ and $d_{14}$, showing contact resistances to be negligible.

Figure 1 reveals that the $Pr_{0.48}Ca_{0.52}MnO_3$ film shows superlattice reflections. We investigated whether the superlattice (with wavevector **q**) is rotated [$\theta$, Fig. 1(a)] with respect to the parent lattice (with wavevector **a***), given that in the bulk there are $\theta$ ~ 0.5° superlattice misorientations that vary on a sub-100 nm length scale, possibly due to coupling with the antiferromagnetic order that develops[11] below $T_N$ ~150 K. For the film, we found $\theta$ = 0 in all 100 nm [010] SAD patterns [e.g. Figs. 1(b)-(d)] measured at all temperatures where superlattice reflections could be detected [Fig. 1(e)]. We could not resolve the possibility of variations in $\theta$ on a shorter length scale, since the acquisition of convergent-beam electron diffraction patterns and dark-field images[11] is precluded by the low intensity of the superlattice reflections. However, it is possible that $\theta$ = 0 locally, as $q/a^*$ = 0.480(2) at 20 K [Fig. 1(e)] matches within error the expected low-temperature value of $(1-x) = 0.48$, unlike bulk[11] $Pr_{0.48}Ca_{0.52}MnO_3$ where $\theta \neq 0$ and $q/a^* \approx 0.46$.

Four-probe d.c. measurements [Fig. 2(a)] show no evidence of the CO transition, unlike bulk samples of similar composition[26], and just like[27] films of similar composition which appear[28] to show no CO. This is consistent with previous observations for (001)-oriented manganite films of other compositions, where the potential feature in resistivity at $T_{CO}$ is weak[29], or even absent in samples where the CO modulation was nevertheless verified by TEM[18,20]. Film resistivity [Fig. 2(a)] also shows no hysteresis associated with antiferromagnetism, unlike bulk samples of similar composition[30]. However, our data reveal that film resistivity is strongly anisotropic. The red line in Fig. 2(a) shows that $\rho_a(T)/\rho_c(T) = 28$ at all measurement temperatures (see later). By contrast, this ratio reaches ~1.1 at 100 K for $La_{0.5}Ca_{0.5}MnO_3$ in ref. 18, but this low value could be due to the irregular contact geometry. Studies of anisotropy in CO manganite films are otherwise scarce. A temperature-independent ratio $\rho_a/\rho_c \approx 2$ was seen for epitaxial films of $La_{0.45}Ca_{0.55}MnO_3$ and $La_{0.33}Ca_{0.67}MnO_3$ on $NdGaO_3$ (001) substrates, but CO was not confirmed by TEM[31].

It is likely that our value for $\rho_a/\rho_c$ is an underestimate, as a small angular misorientation $\phi$ of both electrodes with respect to the crystal axes would dramatically reduce the measured resistivity $\rho$ that is nominally measured along $a$. This may be seen in Fig. 2b (solid line), where we plot the ratio $\rho(\phi)/\rho_c$ via the following formula[32]:

$$\frac{1}{\rho(\phi)} = \frac{\cos^2(\phi)}{\rho_a} + \frac{\sin^2(\phi)}{\rho_c}, \qquad (1)$$

assuming (i) that $\phi = 0$ in our measurement geometry, (ii) that the major axes of the resistivity tensor coincide with $a$ and $c$, and (iii) that $\rho_a/\rho_c = 28$. A misorientation of $\phi = 11°$ would be sufficient to make the two terms on the right equal, and this would halve $\rho(\phi)/\rho_c$. Given this high sensitivity to $\phi$, and given a relative misorientation between the two electrodes of up to ~15°, we anticipate that $\rho_a/\rho_c > 28$. These misorientations, and the possibility that the major axes of the resistivity tensor do not coincide with $a$ and $c$, could explain why $\rho_a/\rho_c$ is temperature independent. The sensitivity to $\phi$ may be avoided by measuring a long thin strip, as seen in Fig. 2b (dashed line) where we use the following formula[32]:

$$\rho(\phi) = \rho_a \cos^2(\phi) + \rho_c \sin^2(\phi), \qquad (2)$$

with the same three assumptions as for Eq. 1. However, this geometry would lead to an unduly high sample resistance and so we did not attempt it.

In Fig. 2(c) we replot the resistivity data from Fig. 2(a) as a function of inverse temperature. Given that $\rho_a(T)/\rho_c(T) = 28$ at all measurement temperatures, the dashed lines for both $a$ and $c$-axis data correspond to the same activation energy (129 meV) for data above ~160 K (~$T_N$). Below this temperature the activation energy falls, but the available temperature range is too narrow to fit the data with confidence.

However, if we assume $\ln(\rho/\rho_0) \propto (T/T_p)^p$ for fit parameters $\rho_0$, $p$ and $T_p$, then a fit with $p = 1/2$ [solid lines, Fig. 2(c)] is marginally better than $p = 1/3$, and the corresponding fit parameter $T_{1/2} = 4.7 \times 10^4$ K is similar to the value ($5.1 \times 10^4$ K) for polycrystalline $Pr_{1/2}Ca_{1/2}MnO_3$ obtained from data over a wider temperature range[25].

The high sensitivity of the current density to angular deviations should not affect the possibility of observing a collective CDW drift due to an electric field $E$. Fig. 3 shows both d.c. and pulsed $J$-$E$ characteristics ($J$ is current density) for measurements of the film at various temperatures along $a$ (upper panel) and $c$ (lower panel). Our key result here is that all d.c. characteristics (solid lines) are ohmic over several orders of magnitude and so there is no evidence for sliding charge density waves along $a$. The non-linear hysteretic regime for $J > 300$-$400$ A/cm$^2$ is attributed to heating[19] as the pulsed data remain virtually linear (10 % deviation at $E = 1000$ V/cm).

In summary, we have used TEM to show that an untwinned (001)-oriented film of $Pr_{0.48}Ca_{0.52}MnO_3$ shows superlattice modulations along $a$. Unlike bulk manganites of this composition[11], the superlattice is aligned with the parent lattice ($\theta = 0$) on a length scale of 100 nm or less. We suggest that this is due to epitaxial strain overcoming a relatively weak[11] local electron-lattice coupling. Failure to observe CDW sliding could be due to strong pinning in our $Pr_{0.48}Ca_{0.52}MnO_3$ film, or a more general failure of manganites and other CO transition-metal oxides to show CDW sliding[19].

Acknowledgements: We are indebted to Dr. K. B. Chashka for his help with the sample contacts in the last two rounds of measurements. We thank J. C. Loudon for investigating $La_{0.5}Ca_{0.5}MnO_3$ films in which no CO superlattice reflections were observed, as noted here. We thank J. Sánchez-Benítez and J. P. Attfield for attempting SQUID magnetometry. This work was supported by an FP6 EU Marie Curie Fellowship (D.S.), UK EPSRC grants EP/E03389X and EP/E0026206. X. M. acknowledges support from Comissionat per a Universitats i Recerca (CUR) del Departament d'Innovació, Universitats i Empresa, de la Generalitat de Catalunya.

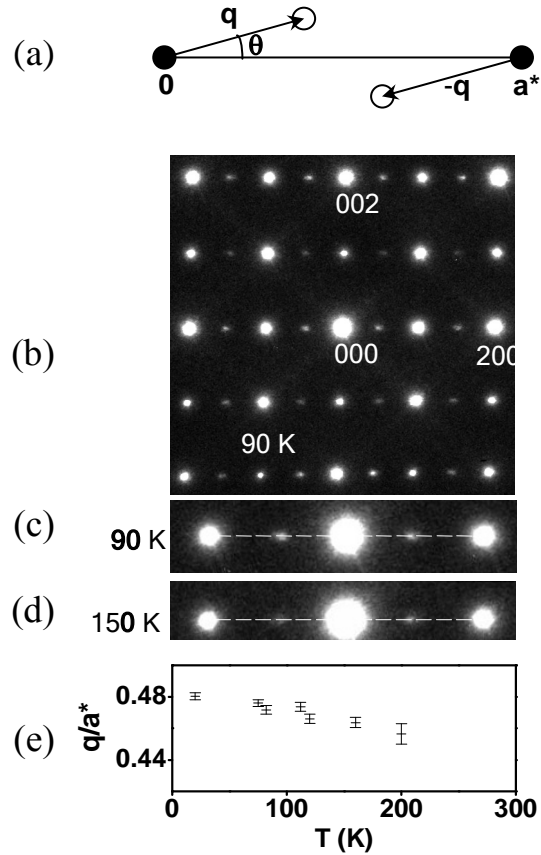

Fig. 1. Electron diffraction data for the 53 nm epitaxial $Pr_{0.48}Ca_{0.52}MnO_3$ film. (a) Schematic showing the relation between parent-lattice reflections (•) and superlattice reflections (○). The possibility of a superlattice misorientation is parameterised via angle $\theta$. (b) A 100 nm [010] SAD pattern taken at 90 K reveals that **q** is parallel to **a***, i.e. $\theta = 0$. Only one of two nearby superlattice reflections is visible, i.e. the one associated with the strongest nearby parent reflection. The origin of the observed diffuse scattering is not resolved. (c) Detail from (b) showing $\theta = 0$ at 90 K, and (d) a corresponding detail showing $\theta = 0$ at 150 K (dashed lines are a guide to the eye). (e) $q/a*$ from 100 nm [010] SAD patterns measured as a function of temperature $T$ on warming. We could not resolve $T_{CO}$ near which the lifetime of superlattice reflections is reduced to a few seconds under electron-beam illumination. Data were collected using a nitrogen stage (b-d) and a helium stage (e). All data were taken in approximately the same electron-transparent region. $\theta = 0$ at all superlattice measurement temperatures.

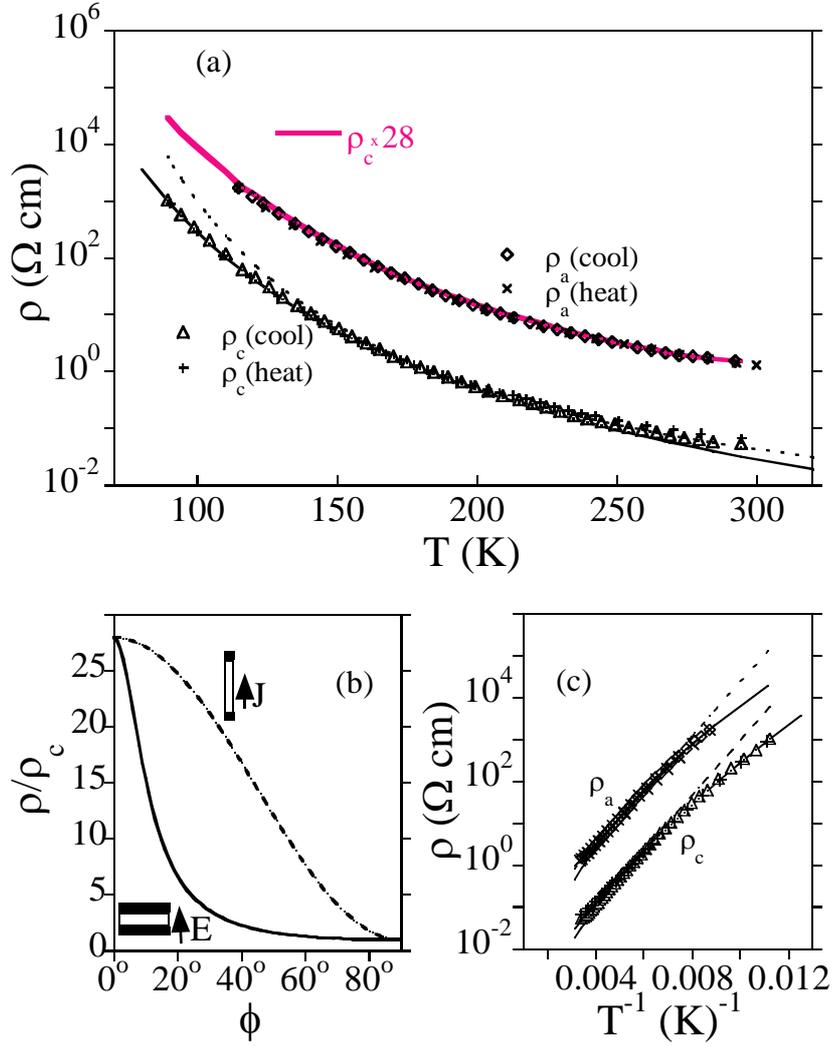

Fig. 2 Four-probe d.c. electrical measurements of the $Pr_{0.48}Ca_{0.52}MnO_3$ film. (a) Semi-log plots of resistivity $\rho$ versus temperature $T$ for cooling and heating measurements along $a$ (upper traces) and $c$ (lower traces). The solid red line represents the fit to the $c$ axis resistivity, multiplied by a factor of 28. (b) Measured resistivity $\rho(\phi)$ normalised by $\rho_c$ as a function of misorientation angle $\phi$ calculated from Eq. 1 (solid line) and Eq. 2 (dashed line) for an anisotropy ratio of 28. The geometries of the electrodes (black) and exposed film (white) are shown schematically next to the corresponding traces. (c) The data from (a) replotted as a function of $T^{-1}$. The straight dashed lines represent activated behaviour, and the solid lines represent the fit described in the text. These fits are also shown in (a) for the $c$-axis dataset (lower traces), but are omitted for the $a$-axis dataset (upper traces) for clarity.

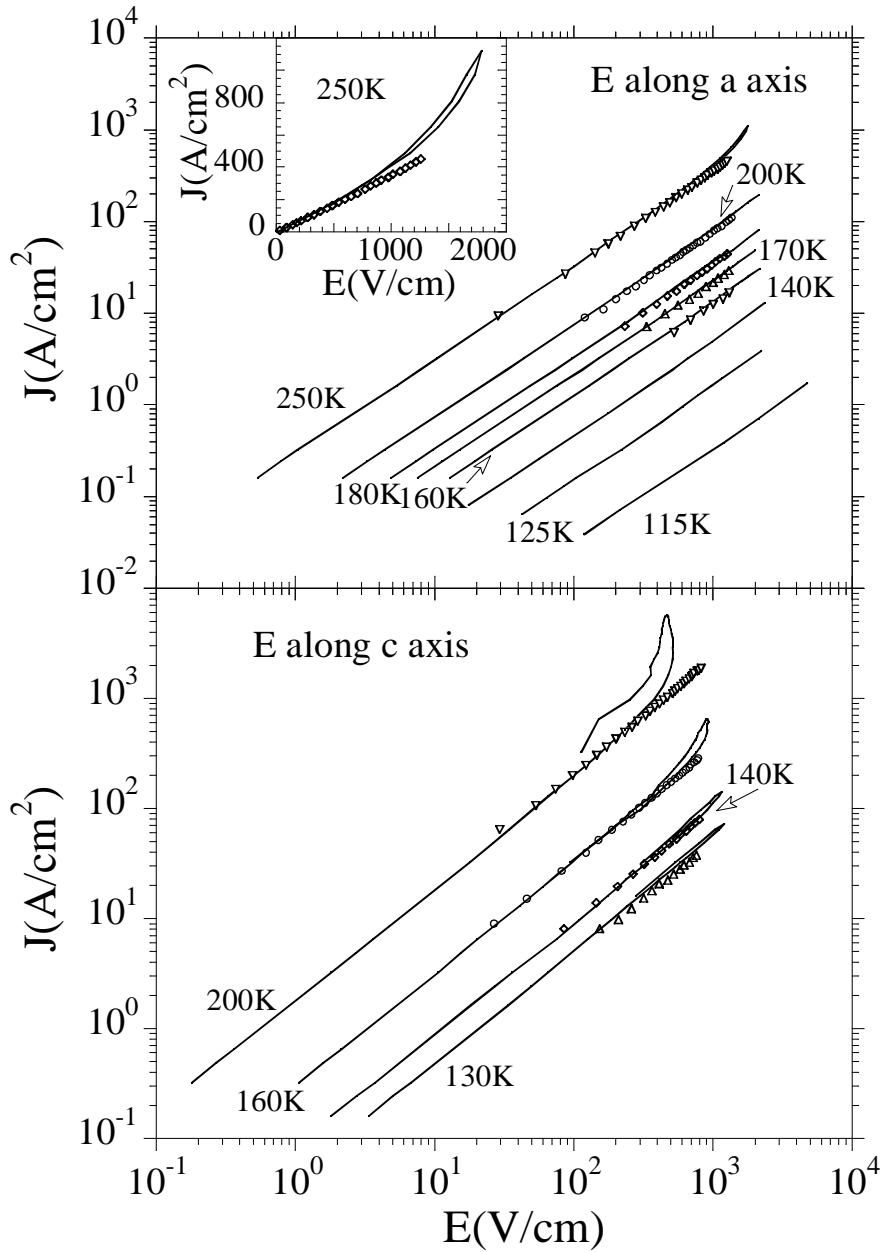

Fig. 3. *J-E* characteristics of the $Pr_{0.48}Ca_{0.52}MnO_3$ film plotted at various temperatures on a log-log scale. Data were measured in a four-probe configuration with *E* along *a* (upper panel) or *c* (lower panel). Both d.c. (solid lines) and pulsed (symbols) measurements are reported. The inset shows the pulsed and d.c. data at 250 K on a linear scale.